\documentclass[aps,prl,twocolumn,amsmath,amssymb,superscriptaddress]{revtex4-1}
\usepackage{dcolumn}
\usepackage{bm}

\usepackage{color} 
\usepackage{ulem}

\usepackage{amsmath}
\usepackage{graphicx}
\usepackage{float}
\usepackage{subfig}
\usepackage{tikz}
\usepackage{color}
\usepackage[colorlinks,bookmarks=false,citecolor=blue,linkcolor=red,urlcolor=blue]{hyperref}

\definecolor{darkred}{rgb}{0.7,0.0,0.0}

\definecolor{darkblue}{rgb}{0,0.02,0.45}

\definecolor{darkgreen}{rgb}{0.02,0.45,0.0}

\definecolor{violet}{rgb}{0.8,0.2,0.6}

\providecommand{\U}[1]{\protect\rule{.1in}{.1in}}

\begin{document}

\title{Exact weak bosonic zero modes in a spin/fermion chain}

\author{Jianlong Fu}
\affiliation{Department of Physics, The University of Hong Kong, Pokfulam Road, Hong Kong, China}

\begin{abstract}
We study an exactly solvable one-dimensional spin-$\frac{1}{2}$ model which can support weak zero modes in its ground state manifold. The spin chain has staggered XXZ-type and ZZ-type spin interaction on neighboring bonds and is thus dubbed the (XXZ,Z) chain. The model is equivalent to an interacting fermionic chain by Jordan-Wigner transformation. We study the phase diagram of the system and work out the conditions and properties of its weak zero modes. In the fermion chain representation, these weak zero modes are given by even-order polynomials of Majorana fermion operators and are thus bosonic. The fermionic chain Hamiltonian contains only fermion hopping and interaction terms and may have potential realization in experiments.
\end{abstract}

\maketitle

The appearance of edge or defect zero modes is the key indication of topological phases, for which a complete classification has been obtained in {\it free} fermionic systems \cite{Chiu16}. Among these zero modes the most fundamental is single Majorana zero modes \cite{alicea20121,beenakker20131}, for which a few superconducting toy models were proposed, including the Kitaev $p$-wave superconducting chain \cite{kitaev01} and some composite models \cite{Fu2021}. In free fermionic systems, the zero modes are given by {\it first-order} fermionic operators that commute with the Hamiltonian itself. According to a classification by Alicea and Fendley \cite{Aliceareview16}, such operators are {\it strong zero modes}. For real systems in which interactions cannot be neglected, strong zero modes are difficult to find and write down explicitly \cite{Fendley2016}, and the results of topological phases for free systems cannot apply directly \cite{Chiu16}. In these systems, more common types of zero modes are {\it weak zero modes}, which are operators that commute with the Hamiltonian projected onto a subspace of the Hilbert space \cite{Aliceareview16,Fendley2012}. From a broader perspective, a weak zero mode is degeneracy of some of the eigenstates of the Hamiltonian, instead of all the eigenstates as for strong zero modes. Although weak zero modes were studied in various systems including the Majorana zero modes associated with vortices and edges in the (bosonic) Kitaev-type chiral spin liquids \cite{Kitaev06,yao2007,Fu20191,peri20} and some one-dimensional systems \cite{wouters18,wada21}, the general nature of such degeneracy is still unclear; for example it is interesting to ask whether the statistics of weak zero modes in fermionic systems is necessarily fermionic, and how the weak zero modes are localized. 

In this work we study an exactly solvable one-dimensional spin-$\frac{1}{2}$ chain which has an equivalent interacting fermion-chain representation by Jordan-Wigner (JW) transformation \cite{Jordan1928,Lieb61}. The model is solved exactly using another Jordan-Wigner transformation in a rotated basis \cite{sachdevbook,Mcginley2017}. By examining its physical states, which are described by fermions coupled to static $Z_{2}$ variables, we show that ground-state weak zero modes appear on the edges of the chain in some phases of the model. Such weak zero modes can be exactly written down in both the spin-chain and fermion-chain representations. Moreover, the fermion-chain representation is a simple interacting model; compared with previously studied models for weak zero modes \cite{wouters18,wada21,ort14}, in which sometimes superconducting terms and fermion interaction terms coexist in the Hamiltonian, the simplicity of our model facilitates potential {\it realizations in experiments}. The weak zero modes in the fermion chain are given by {\it polynomials} of Majorana operators and thus represent a {\it particle-hole type} degeneracy in certain {\it subspaces} of the physical Hilbert space \cite{Mcginley2017}. Depending on the order of the polynomials, these weak zero modes are not necessarily fermionic; in our case the weak zero modes are {\it bosonic} since the polynomials contain only even order terms. Many previous proposals of exactly-solvable or quasi-exactly-solvable one-dimensional (1D) {\it interacting} models have fermionic zero modes \cite{Mcginley2017,sau11,Fid12,ort14}. On the contrary, our results show that weak zero modes in fermionic systems can be bosonic. As can be seen later, the construction of our model can generalize to other models containing weak zero modes. 
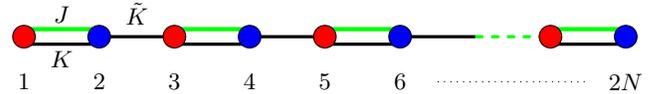
\begin{figure}
	\begin{tikzpicture}
	\foreach \i in {0,2,4}
	{
		\draw[very thick](\i,-0.1)--(\i+1,-0.1);
		\draw[ultra thick,green](\i,0.1)--(\i+1,0.1);
		\draw[very thick](\i+1,0)--(\i+2,0);
		\filldraw[fill=red] (\i,0) circle [radius=0.15];
		\filldraw[fill=blue] (\i+1,0) circle [radius=0.15]; 
	}
	\foreach \j in {1,2,3,4,5,6}
	{
		\node at (\j-1,-0.6) {\j};
	}
	\node at (0.5,-0.3) {$K$};
	\node at (0.5,0.3) {$J$};
	\node at (1.5,0.3) {$\tilde{K}$};
	\draw[very thick,dashed,green](6,0)--(7,0);
	\draw[very thick](7,-0.1)--(8,-0.1);
	\draw[ultra thick,green](7,0.1)--(8,0.1);
	\filldraw[fill=red] (7,0) circle [radius=0.15];
	\filldraw[fill=blue] (8,0) circle [radius=0.15];
	\node at (8,-0.6) {$2N$};
	\draw[dotted](5.5,-0.6)--(7.5,-0.6);
	\end{tikzpicture}
	\caption{The lattice of the (XXZ, Z) spin-chain model. The black bonds indicate the ZZ spin coupling and the green bonds are the XY spin coupling. For the fermionic representation of the model the black bonds indicate the fermion interaction terms and the green bonds represent the fermion hopping terms.}
	\label{figlattice}
\end{figure}

\section{The model and its solution}

To introduce the model, we start with a spin-$\frac{1}{2}$ chain of finite length, the spins are labeled by integer $n$ which goes from $1$ to $2N$ ($N$ is a large integer).  The spins interact with their nearest neighbours by a staggered $XXZ$ and $ZZ$ coupling, one unit cell of the model consists of two sites $(2n-1,2n)$. The Hamiltonian of the spin chain is given by
\begin{eqnarray}
\begin{aligned}
\label{spinchainH}
\mathcal{H}=&\sum_{n=1}^{N}\bigg[J_{n}\big(\sigma_{2n-1}^{x}\sigma_{2n}^{x}+\sigma_{2n-1}^{y}\sigma_{2n}^{y}\big)+K_{n}\sigma_{2n-1}^{z}\sigma_{2n}^{z}\bigg]\\&+\sum_{m=1}^{N-1}\tilde{K}_{m}\sigma_{2m}^{z}\sigma_{2m+1}^{z},
\end{aligned}
\end{eqnarray}
in which $J_{n}$ and $K_{n}$ give the $XXZ$ coupling strength on bonds $(2n-1,2n)$ and $\tilde{K}_{m}$ gives $ZZ$ coupling strength on bonds $(2m,2m+1)$, as shown in Fig. \ref{figlattice}. The model is thus referred to as the {\it (XXZ, Z) spin chain}. It is easily seen that the spin Hamiltonian (\ref{spinchainH}) has conserved Ising bilinears in every unit cell, namely $[\sigma_{2n-1}^{z}\sigma_{2n}^{z},\mathcal{H}]=0$ for all $n$. These conserved variables imply that the Hilbert space is divided into independent Krylov subspaces, the number of which scales linearly with the size of the system \cite{moudgalya22}.

The spin chain model (\ref{spinchainH}) can be transformed into a fermionic chain by Jordan-Wigner transformation \cite{Jordan1928,Lieb61}. To this end we introduce a fermion $c_{i}^{\dagger}$ for every spin $\boldsymbol{\sigma}_{i}$ and require $\sigma_{i}^{+}=c_{i}^{\dagger}(-1)^{\sum_{j=1}^{i-1}c_{j}^{\dagger}c_{j}}$ as well as $\sigma_{i}^{z}=2c_{i}^{\dagger}c_{i}-1$, under which the original spin model (\ref{spinchainH}) becomes
\begin{eqnarray}
\begin{aligned}
\label{fermionicH}
\mathcal{H}=&\sum_{n=1}^{N}\bigg[2J_{n}\big(c_{2n-1}^{\dagger}c_{2n}+c_{2n}^{\dagger}c_{2n-1}\big)\\&+K_{n}\big(2c_{2n-1}^{\dagger}c_{2n-1}-1\big)\big(2c_{2n}^{\dagger}c_{2n}-1\big)\bigg]\\&+\sum_{m=1}^{N-1}\tilde{K}_{m}\big(2c_{2m}^{\dagger}c_{2m}-1\big)\big(2c_{2m+1}^{\dagger}c_{2m+1}-1\big).
\end{aligned}
\end{eqnarray}
This is a simple interacting fermionic Hamiltonian, its kinetic hopping terms vanish for every second bond (see Fig. \ref{figlattice}) and the fermions interact by their density fluctuations, possibly with respect to a uniform positive charge background (of $\frac{1}{2}$ per site). Experimentally the fermionic model can possibly be realized in Su-Schrieffer-Heeger systems \cite{Su791} with Peierls instability, in which the translational symmetry is broken such that fermion hopping can be neglected for every second bond although interaction between neighboring sites cannot be neglected. 

\begin{figure}
	\begin{tikzpicture}
	\node at (-0.8,0) {$\eta^{\beta}$};
	\node at (-0.8,0.6) {$\eta^{\alpha}$};
	\foreach \i in {1,2,3,4,5,6}
	{
		\node at (\i*0.8-0.8,-0.5) {\i};
	}
	\foreach \j in{0,2,4}
	{
		\draw[very thick,dashed,green](\j*0.8,0.6)--(\j*0.8+0.8,0);
		\draw[very thick,orange](\j*0.8,0)--(\j*0.8+0.8,0.6);
	}
	\foreach \j in{1,3}
	{
		\draw[very thick](\j*0.8,0.6)--(\j*0.8+0.8,0);
	}
	\foreach \i in {0,1,2,3,4,5}
	{
		\filldraw[fill=red] (\i*0.8,0) circle [radius=0.08];
		\filldraw[fill=blue] (\i*0.8,0.6) circle [radius=0.08];
	}
	\draw[very thick,dotted](4.5,0.3)--(5.1,0.3);
	
	\draw[very thick,dashed,green](5.6,0.6)--(6.4,0);
	\draw[very thick,orange](5.6,0)--(6.4,0.6);
	
	\foreach \i in {7,8}
	{
		\filldraw[fill=red] (\i*0.8,0) circle [radius=0.08];
		\filldraw[fill=blue] (\i*0.8,0.6) circle [radius=0.08];
	}
	\node at (5.6,-0.5) {2N-1};
	\node at (6.4,-0.5) {2N};
	\draw[dotted](4.3,-0.5)--(5.1,-0.5);
	\end{tikzpicture}
	\caption{The model is solved by a rotated Jordan-Wigner transformation. Majorana bilinears that commute with the Hamiltonian are denoted by green dashed lines, which in turn form the static $Z_{2}$ variables. The solution is a Majorana hopping model, connected by orange and black lines, coupled with static $Z_{2}$ variables.}
	\label{figeffective}
\end{figure}
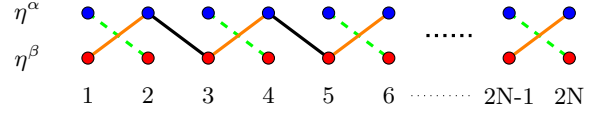

The spin chain Hamiltonian (\ref{spinchainH}) and the fermionic Hamiltonian (\ref{fermionicH}) form the two {\it representations} of the same model. The two representations enjoy a one-to-one correspondence between their Hilbert space and eigenstates, hence the physical properties such as degeneracies are identical. The model (\ref{spinchainH}) can be solved by another Jordan-Wigner transformation in a rotated basis $(x,y,z)\rightarrow (y,z,x)$ \cite{sachdevbook,Mcginley2017}; to this end we introduce another set of fermions $d_{i}$ and define $\sigma_{i}^{y}=(d_{i}+d_{i}^{\dagger})(-1)^{\sum_{j=1}^{i-1}d_{j}^{\dagger}d_{j}}$, $\sigma_{i}^{z}=i(d_{i}-d_{i}^{\dagger})(-1)^{\sum_{j=1}^{i-1}d_{j}^{\dagger}d_{j}}$ as well as $\sigma_{i}^{x}=2d_{i}^{\dagger}d_{i}-1$. It simplifies the problem to decouple the $d_{i}$ fermions into Majorana fermions $\eta_{i}^{\alpha}$ and $\eta_{i}^{\beta}$ by $d_{i}^{\dagger}=\frac{1}{2}(\eta_{i}^{\alpha}+i\eta_{i}^{\beta})$. In terms of the Majorana fermions the new Jordan-Wigner transformation can be written as
\begin{eqnarray}
\label{rotatedJW}
\begin{aligned}
\sigma_{i}^{x}&=-i\eta_{i}^{\alpha}\eta_{i}^{\beta},\qquad \sigma_{i}^{y}=\eta_{i}^{\alpha}\prod_{j=1}^{i-1}(i\eta_{j}^{\alpha}\eta_{j}^{\beta}),\\ \sigma_{i}^{z}&=\eta_{i}^{\beta}\prod_{j=1}^{i-1}(i\eta_{j}^{\alpha}\eta_{j}^{\beta}).
\end{aligned}
\end{eqnarray}
Using the transformation (\ref{rotatedJW}), the original spin model (\ref{spinchainH}) can be written as
\begin{eqnarray}
\label{MajoranaH}
\begin{aligned}
\mathcal{H}=\sum_{n=1}^{N}&\bigg[J_{n}(i\eta_{2n-1}^{\alpha}\eta_{2n}^{\beta})(i\eta_{2n-1}^{\beta}\eta_{2n}^{\alpha})+J_{n}i\eta_{2n-1}^{\beta}\eta_{2n}^{\alpha}\\&-K_{n}i\eta_{2n-1}^{\alpha}\eta_{2n}^{\beta}\bigg]+\sum_{m=1}^{N-1}\tilde{K}_{m}(-i)\eta_{2m}^{\alpha}\eta_{2m+1}^{\beta}.
\end{aligned}
\end{eqnarray}
In this Majorana Hamiltonian we notice that the Majorana bilinears $i\eta_{2n-1}^{\alpha}\eta_{2n}^{\beta}$ commute with the Hamiltonian; therefore a static $Z_{2}$ variable $\tau_{n}=i\eta_{2n-1}^{\alpha}\eta_{2n}^{\beta}$ can be introduced for every unit cell corresponding to the conserved Ising bond-operators in the spin representation. The Majorana Hamiltonian is illustrated in Fig. \ref{figeffective}. With these definitions, the Hamiltonian is finally written as
\begin{eqnarray}
\label{HMajorana}
\begin{aligned}
\mathcal{H}=&\sum_{n=1}^{N}\bigg[J_{n}(1+\tau_{n})i\eta_{2n-1}^{\beta}\eta_{2n}^{\alpha}-K_{n}\tau_{n}\bigg]\\&+\sum_{m=1}^{N-1}\tilde{K}_{m}(-i)\eta_{2m}^{\alpha}\eta_{2m+1}^{\beta},
\end{aligned}
\end{eqnarray}
which takes the form of a Majorana hopping model coupled with static $Z_{2}$ variables. To determine the physical eigenstates of the model, one first takes a set of $\{\tau_{n}\}$, under which the Hamiltonian becomes a Majorana hopping model; fermion eigenstates can then be worked out accordingly, and the physical eigenstates of the model take the form of $|\psi\rangle_{E}=|\tau_{n}\rangle\otimes|\eta^{\alpha,\beta}_{E}\rangle_{\{\tau\}}$. Every distribution of the $Z_{2}$ variables and its corresponding fermionic space can be understood as an {\it invariant sector} (Krylov subspace) of the Hilbert space. All physical eigenstates of the model can be worked out sector by sector. The model can also be solved using the SO(3) Majorana representation by introducing three Majorana fermions for each spin, which is expected because of the equivalence between Jordan-Wigner transformation and the SO(3) Majorana representation \cite{berezin1975,Fu20181,Fu20182,Fu20191}.

\section{Phase diagram}

The model (\ref{MajoranaH}) is invariant under all the $J_{n}$ and $\tilde{K}_{n}$ changing signs, accompanied by $\eta_{2m+1}^{\beta}\rightarrow -\eta_{2m+1}^{\beta}$ for all $m$. In light of this symmetry, to simplify further treatment of the Hamiltonian (\ref{HMajorana}) we take all the parameters $\tilde{K}_{m}\equiv -1$ and assume that all $J_{n}$ and $K_{n}$ are constants, namely $J_{n}\equiv J$ and $K_{n}\equiv K$. Among all the physical eigenstates of the model, we are most interested in the ground state.  To look for the ground state, it is noteworthy that the solution of our model shares some similarities with the Jordan-Wigner transformation solution of the Kitaev honeycomb model \cite{Feng07,schmidt08}; there the static $Z_{2}$ variables are located on one of the three types of bonds. Such similarity allows us to borrow the results from the Kitaev honeycomb model and make the {\it assumption} that the ground state lies in the sectors in which all $\tau_{n}$ are equal \cite{Kitaev06,lieb94}. In other words, the ground state is searched for in two sectors $\{\tau_{n}\equiv 1\}$ and $\{\tau_{n}\equiv -1\}$. The sector that contains the ground state is then referred to as the {\it ground-state sector}. 

Under these assumptions the Hamiltonian (\ref{HMajorana}) becomes 
\begin{equation}
\label{Hsimplified}
\mathcal{H}=\sum_{n=1}^{N}\bigg[J(1+\tau)i\eta_{2n-1}^{\beta}\eta_{2n}^{\alpha}-K\tau\bigg]+\sum_{m=1}^{N-1}i\eta_{2m}^{\alpha}\eta_{2m+1}^{\beta}.
\end{equation}  
To find the ground state energy, we pair up $\eta_{2n-1}^{\beta}$ and $\eta_{2n}^{\alpha}$ in each unit cell to define a complex fermion $f_{n}^{\dagger}=\frac{1}{2}\big(\eta_{2n-1}^{\beta}+i\eta_{2n}^{\alpha}\big)$. Due to the {\it translational invariance} of the Hamilonian (\ref{Hsimplified}), a Fourier transformation can be performed, $f_{n}^{\dagger}=\frac{1}{\sqrt{N}}\sum_{k}f_{k}^{\dagger}e^{ikn}$. In momentum space, the Hamiltonian (\ref{Hsimplified}) is given by
\begin{equation}
\mathcal{H}=-NK\tau+\sum_{k}\left(\begin{array}{cc}
f_{k}^{\dagger}&f_{-k}
\end{array}\right)\hat{\mathbf{H}}_{k}\left(\begin{array}{c}
f_{k}\\f_{-k}^{\dagger}
\end{array}\right),
\end{equation}
in which 
\begin{equation}
\hat{\mathbf{H}}_{k}=\left(\begin{array}{cc}
\cos k-J(1+\tau)&-i\sin k\\
i\sin k&J(1+\tau)-\cos k
\end{array}\right).
\end{equation}
The eigenvalues of the matrix $\hat{\mathbf{H}}_{k}$ are $E_{k}=\pm\sqrt{1+J^{2}(1+\tau)^{2}-2J(1+\tau)\cos k}$. As the number of unit cells $N\rightarrow \infty$, the ground state energy density is given by the integral 
\begin{equation}
\frac{E_{0}}{N}=-\frac{1}{2\pi}\int_{-\pi}^{\pi}dk \sqrt{1+4J^{2}-4J\cos k}-K
\end{equation}
for the $\tau\equiv 1$ sector, and 
\begin{equation}
\frac{E_{0}}{N}=K-1
\end{equation}
for the $\tau\equiv -1$ sector. To determine which sector has the lowest ground-state energy one needs to compare these two values for each $J$ and $K$. 

\begin{figure}
\includegraphics[width=0.4\textwidth]{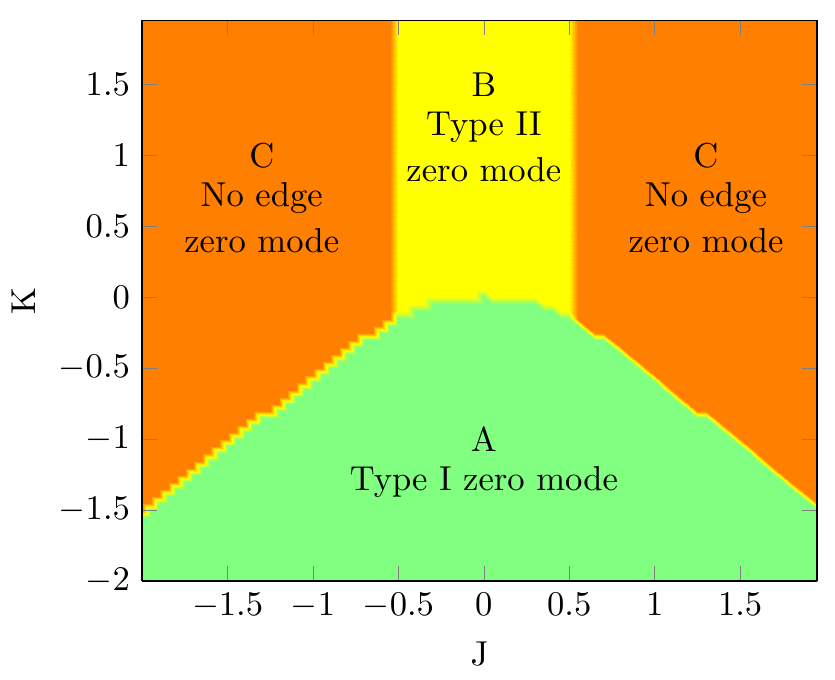}
\caption{Phase diagram of the model with respect to coupling $J$ and $K$. Phase A has type I edge zero modes. Phase B has type II edge zero modes and phase C has no edge zero modes.}
\label{figphase}
\end{figure}

We now turn to discuss the possible zero modes in the Majorana hopping model in every sector. To do so, it helps to consider the chain as located on a closed circle. As can be seen from the Hamiltonian (\ref{HMajorana}) the edge corresponds to a single broken $\tilde{K}$ bond. The ground-state sector can only have {\it edge zero modes} since we assumed that it does not break translational invariance. For the excited states sectors, the situation varies according to the distribution of $\tau_{n}$; for some of the distributions the chain can have broken $J-K$ bonds (when $\tau_{n}=-1$ on that bond), causing defects for the corresponding Majorana hopping model. In these excited-state-sectors, the possible Majorana zero modes are complicated, but can still be exactly written down. We have the following results on its edge zero modes for the translationally invariant sector, and defect zero modes in the excited-states sectors. 

(i) If the ground state of the model is in the sector $\tau_{n}\equiv -1$, then we have {\it strictly localized} Majorana zero modes $\eta_{1}^{\beta}$ and $\eta_{2N}^{\alpha}$ on both edges of the chain. As can be seen in the Hamiltonian (\ref{Hsimplified}), this case is effectively equivalent to $J=0$. We call this type of edge zero modes {\it type I} zero modes and such phase is referred to as {\it phase A}. The ground state of the model is degenerate in phase A.

(ii) If the ground state of the model is in the sector $\tau_{n}\equiv 1$, the Majorana hopping model from the Hamiltonian (\ref{Hsimplified}) becomes a {\it first-order model} as defined in Ref. \cite{Fu2021}, and we have two situations for its Majorana zero modes. (a) When $|J|<\frac{1}{2}$, there are {\it localized} edge Majorana zero modes on both ends whose wavefunction is exponentially decaying into the system \cite{Fu2021}. We refer to this kind of edge zero modes as {\it type II} zero modes and the phase as {\it phase B}. For excited states in other sectors without translational invariance, some of the bonds have $\tau_{n}=-1$, but no defect zero modes are associated with these bonds. (b) When $|J|>\frac{1}{2}$, there is no localized edge zero modes in the ground-state sector. However, localized defect zero modes exist around $\tau_{n}^{a}=-1$ bonds in other excited-state sectors, provided that these defects are well-separated from each other. This phase is called {\it phase C}. The ground state is not degenerate in phase C.

Unlike type I zero modes, whose existence implies {\it exact degeneracy} among the states within its sector, type II zero modes and other defect zero modes whose wave functions are not strictly localized means that the states are only {\it approximately} degenerate. For all sectors, the split between (potentially degenerate) energy levels, which exponentially decays with the distances between these zero modes, can only be taken as vanishing when the $\tau_{n}=-1$ bonds are well separated from each other. It is thus impossible to write down a universal operator that commutes with the original Hamiltonian (\ref{HMajorana}) and brings degeneracy in all sectors. Therefore both type I and type II edge zero modes are {\it weak zero modes} \cite{Aliceareview16,Fendley2012}, which only imply degeneracy in certain subspaces (or sectors) of the Hilbert space. After numerically evaluating the ground-state energies in the two translationally invariant sectors, we arrive at the phase diagram Fig. \ref{figphase}. 

\section{Weak bosonic zero modes}

To understand the nature of the two types of zero modes, we consider a simple XXZ model of a two-spin system, whose Hamiltonian is $\mathcal{H}=J(\sigma_{1}^{x}\sigma_{2}^{x}+\sigma_{1}^{y}\sigma_{2}^{y})+K\sigma_{1}^{z}\sigma_{2}^{z}$. It has two degenerate eigenstates $|\uparrow\uparrow\rangle$ and $|\downarrow\downarrow\rangle$, with energy $K$; the other two eigenstates of the model are $\frac{1}{\sqrt{2}}(|\uparrow\downarrow\rangle+|\downarrow\uparrow\rangle)$, with energy $2J-K$, and $\frac{1}{\sqrt{2}}(|\uparrow\downarrow\rangle-|\downarrow\uparrow\rangle)$, with energy $-2J-K$. The pair of degenerate states with energy $K$ corresponds to the type I zero mode in our original model (\ref{spinchainH}), which is strictly localized in $d$-fermion language but strictly {\it non-localized} in spin language and $c$-fermion language. Moreover, the degenerate ground states of the original spin model are $|\uparrow\uparrow\uparrow\cdots\uparrow\rangle$ and $|\downarrow\downarrow\downarrow\cdots\downarrow\rangle$ in phase A, because we have translations from Majorana fermions back to the spin operators $\eta_{i}^{\alpha}=\sigma_{i}^{y}\prod_{j=1}^{i-1}(-\sigma_{j}^{x})$ and $\eta_{i}^{\beta}=\sigma_{i}^{z}\prod_{j=1}^{i-1}(-\sigma_{j}^{x})$. Specifically for type I Majorana zero modes in phase A we have $\eta_{1}^{\beta}\rightarrow \sigma_{1}^{z}$ and $\eta_{2N}^{\alpha}\rightarrow i\sigma_{2N}^{z}\prod_{j=1}^{2N}\sigma_{j}^{x}$, which give the zero-mode operators in spin language. Although the product operator $\prod\sigma_{i}^{x}$ commutes with the spin Hamiltonian (\ref{spinchainH}), this does not lead to strong zero modes just like in the two-spin XXZ model. As for type II zero modes in phase B, we focus on the two non-degenerate states in the two-spin model. Upon enlarging the length of the spin chain, these two states evolve into a spin-wave-like {\it band} of eigenstates and the type II zero modes exist in this band. The condition for the existence of the type II zero modes is exactly given by the topological condition of the Kitaev chain \cite{kitaev01,Fu2021,Niu12}, such a condition distinguishes phase B and phase C.   

To go to the $c$-fermion language for the fermionic representation of the model (\ref{fermionicH}), we first decouple the $c$ fermion into Majorana fermions, $c_{i}=\frac{1}{2}(\gamma_{i}^{\alpha}-i\gamma_{i}^{\beta})$. The transformation between the two types of Majoranas can be obtained by noting that $\sigma_{i}^{z}=i\gamma_{i}^{\beta}\gamma_{i}^{\alpha}$, as well as $\eta_{i}^{\alpha}=\gamma_{i}^{\alpha}\gamma_{i}^{\beta}\prod_{j=1}^{i}(-\sigma_{j}^{x})$ and $\eta_{i}^{\beta}=i\gamma_{i}^{\beta}\gamma_{i}^{\alpha}\prod_{j=1}^{i-1}(-\sigma_{j}^{x})$. In general the transformation between $\eta$ and $\gamma$ Majorana fermions can be worked out. To this end we have $\eta_{n}^{\alpha}$ is a product of $n$ $\gamma$ Majorana fermions, namely,
\begin{eqnarray}
\begin{aligned}
\eta_{2m}^{\alpha}&=-(i)^{m}\gamma_{1}^{\beta}\gamma_{2}^{\alpha}\cdots\gamma_{2m-1}^{\beta}\gamma_{2m}^{\beta},\\ \eta_{2m+1}^{\alpha}&=(i)^{m}\gamma_{1}^{\alpha}\gamma_{2}^{\beta}\cdots\gamma_{2m}^{\beta}\gamma_{2m+1}^{\beta}.
\end{aligned}
\end{eqnarray}
In addition, $\eta_{n}^{\beta}$ is a product of $n+1$ $\gamma$ Majorana fermions,
\begin{eqnarray}
\begin{aligned}
\eta_{2m}^{\beta}&=-(i)^{m}\gamma_{1}^{\alpha}\gamma_{2}^{\beta}\cdots\gamma_{2m-2}^{\beta}\gamma_{2m-1}^{\alpha}\gamma_{2m}^{\beta}\gamma_{2m}^{\alpha},\\ \eta_{2m+1}^{\beta}&=(i)^{m+1}\gamma_{1}^{\beta}\gamma_{2}^{\alpha}\cdots\gamma_{2m}^{\alpha}\gamma_{2m+1}^{\beta}\gamma_{2m+1}^{\alpha}.
\end{aligned}
\end{eqnarray}
Therefore, the type I zero mode, which is a product of $\gamma$ Majorana fermions from all sites, is strictly {\it non-localized}. The type-II zero modes whose wave function is exponentially decaying into the system are given by $\zeta^{\beta}_{\text{L}}=\sum_{n=1}^{N}\tilde{\lambda}_{2n-1}\eta_{2n-1}^{\beta}$, which is localized on the left end of the chain, and $\zeta^{\alpha}_{\text{R}}=\sum_{n=N}^{1}\tilde{\lambda}'_{2n}\eta_{2n}^{\alpha}$, which is localized on the right end of the chain. They are both translated into polynomials of $\gamma$ Majorana fermions \cite{Mcginley2017}; they imply that the states of the projected subset of the Hilbert space are paired up into degenerate pairs, for every pair the fermion occupation of one state is related to the other by the Majorana polynomial. In our model the Majorana polynomials on both ends involve only even-order terms of $\gamma$ Majorana fermions, indicating that these zero modes are {\it bosonic} in the $c$-fermion language. In other words, the two degenerate states connected by the weak zero mode have the same fermion parity. This is understandable because the static $Z_{2}$ variables $\tau_{n}$ are given by fermion parity in each unit cell $\tau_{n}=-(-1)^{n_{2n-1}+n_{2n}}$ in the $c$-fermion language, with $n_{i}$ being the occupation number on site $i$; these static $Z_{2}$ variables classify the subset of the Hilbert space, thus the weak zero modes in each subset must commute with the $\tau_{n}$ operators and fermion parity. Also notice that the polynomials involve terms that are macroscopic order of $\gamma$ Majorana fermions so that the locality of the weak zero modes has some subtleties \cite{Greiter14}.

\section{Conclusion}

To summarize, we studied an exactly solvable spin chain which has a dual description of interacting fermions in one dimension. The model possesses weak bosonic zero modes in some sectors of the Hilbert space, which can potentially be probed in photo-absorption experiments. Future studies on this model can generalize the discussion beyond the translational invariance assumption and consider possible defect-induced zero modes. Moreover, influences of other terms that bring the Hamiltonian away from the exactly solvable point are interesting to explore; in particular, adding Zeeman terms ($\sum_{n}h_{n}\sigma_{n}^{z}$) in the original spin model (\ref{spinchainH}) preserves the conserved Ising bond-operators and thus is worth considering. The duality transformation between Majorana and bosonic polynomial zero modes is possible only for one-dimensional systems thanks to the JW transformation between bosonic and fermionic degrees of freedom. Nevertheless, the results unveil some properties of weak zero modes in real systems in which interactions are included; in particular the solution exemplifies that the weak zero modes in interacting systems can be {\it non-local} and also {\it bosonic} despite being a model of fermions. Looking forward, the structure of the model can be used to construct other models possessing weak zero modes. The first step is enlarging the Hilbert space by bringing in extra static degrees of freedom, which are then coupled to the original (free) topological model as {\it tuning parameters}. In this way different phases of the original free model can potentially be realized in the same model.

\section*{Acknowledgements}

The author thanks H. C. Po and J. Knolle for insightful discussion and the University of Hong Kong for its support.


\bibliography{refJFU}

\begin{thebibliography}{31}%
\makeatletter
\providecommand \@ifxundefined [1]{%
 \@ifx{#1\undefined}
}%
\providecommand \@ifnum [1]{%
 \ifnum #1\expandafter \@firstoftwo
 \else \expandafter \@secondoftwo
 \fi
}%
\providecommand \@ifx [1]{%
 \ifx #1\expandafter \@firstoftwo
 \else \expandafter \@secondoftwo
 \fi
}%
\providecommand \natexlab [1]{#1}%
\providecommand \enquote  [1]{``#1''}%
\providecommand \bibnamefont  [1]{#1}%
\providecommand \bibfnamefont [1]{#1}%
\providecommand \citenamefont [1]{#1}%
\providecommand \href@noop [0]{\@secondoftwo}%
\providecommand \href [0]{\begingroup \@sanitize@url \@href}%
\providecommand \@href[1]{\@@startlink{#1}\@@href}%
\providecommand \@@href[1]{\endgroup#1\@@endlink}%
\providecommand \@sanitize@url [0]{\catcode `\\12\catcode `\$12\catcode
  `\&12\catcode `\#12\catcode `\^12\catcode `\_12\catcode `\%12\relax}%
\providecommand \@@startlink[1]{}%
\providecommand \@@endlink[0]{}%
\providecommand \url  [0]{\begingroup\@sanitize@url \@url }%
\providecommand \@url [1]{\endgroup\@href {#1}{\urlprefix }}%
\providecommand \urlprefix  [0]{URL }%
\providecommand \Eprint [0]{\href }%
\providecommand \doibase [0]{http://dx.doi.org/}%
\providecommand \selectlanguage [0]{\@gobble}%
\providecommand \bibinfo  [0]{\@secondoftwo}%
\providecommand \bibfield  [0]{\@secondoftwo}%
\providecommand \translation [1]{[#1]}%
\providecommand \BibitemOpen [0]{}%
\providecommand \bibitemStop [0]{}%
\providecommand \bibitemNoStop [0]{.\EOS\space}%
\providecommand \EOS [0]{\spacefactor3000\relax}%
\providecommand \BibitemShut  [1]{\csname bibitem#1\endcsname}%
\let\auto@bib@innerbib\@empty
\bibitem [{\citenamefont {Chiu}\ \emph {et~al.}(2016)\citenamefont {Chiu},
  \citenamefont {Teo}, \citenamefont {Schnyder},\ and\ \citenamefont
  {Ryu}}]{Chiu16}%
  \BibitemOpen
  \bibfield  {author} {\bibinfo {author} {\bibfnamefont {C.-K.}\ \bibnamefont
  {Chiu}}, \bibinfo {author} {\bibfnamefont {J.~C.~Y.}\ \bibnamefont {Teo}},
  \bibinfo {author} {\bibfnamefont {A.~P.}\ \bibnamefont {Schnyder}}, \ and\
  \bibinfo {author} {\bibfnamefont {S.}~\bibnamefont {Ryu}},\ }\href {\doibase
  10.1103/RevModPhys.88.035005} {\bibfield  {journal} {\bibinfo  {journal}
  {Rev. Mod. Phys.}\ }\textbf {\bibinfo {volume} {88}},\ \bibinfo {pages}
  {035005} (\bibinfo {year} {2016})}\BibitemShut {NoStop}%
\bibitem [{\citenamefont {Alicea}(2012)}]{alicea20121}%
  \BibitemOpen
  \bibfield  {author} {\bibinfo {author} {\bibfnamefont {J.}~\bibnamefont
  {Alicea}},\ }\href {\doibase 10.1088/0034-4885/75/7/076501} {\bibfield
  {journal} {\bibinfo  {journal} {Reports on Progress in Physics}\ }\textbf
  {\bibinfo {volume} {75}},\ \bibinfo {pages} {076501} (\bibinfo {year}
  {2012})}\BibitemShut {NoStop}%
\bibitem [{\citenamefont {Beenakker}(2013)}]{beenakker20131}%
  \BibitemOpen
  \bibfield  {author} {\bibinfo {author} {\bibfnamefont {C.}~\bibnamefont
  {Beenakker}},\ }\href {\doibase 10.1146/annurev-conmatphys-030212-184337}
  {\bibfield  {journal} {\bibinfo  {journal} {Annual Review of Condensed Matter
  Physics}\ }\textbf {\bibinfo {volume} {4}},\ \bibinfo {pages} {113} (\bibinfo
  {year} {2013})}\BibitemShut {NoStop}%
\bibitem [{\citenamefont {Kitaev}(2001)}]{kitaev01}%
  \BibitemOpen
  \bibfield  {author} {\bibinfo {author} {\bibfnamefont {A.~Y.}\ \bibnamefont
  {Kitaev}},\ }\href {http://stacks.iop.org/1063-7869/44/i=10S/a=S29}
  {\bibfield  {journal} {\bibinfo  {journal} {Physics-Uspekhi}\ }\textbf
  {\bibinfo {volume} {44}},\ \bibinfo {pages} {131} (\bibinfo {year}
  {2001})}\BibitemShut {NoStop}%
\bibitem [{\citenamefont {Fu}(2021)}]{Fu2021}%
  \BibitemOpen
  \bibfield  {author} {\bibinfo {author} {\bibfnamefont {J.}~\bibnamefont
  {Fu}},\ }\href {\doibase https://doi.org/10.1016/j.aop.2021.168564}
  {\bibfield  {journal} {\bibinfo  {journal} {Annals of Physics}\ }\textbf
  {\bibinfo {volume} {432}},\ \bibinfo {pages} {168564} (\bibinfo {year}
  {2021})}\BibitemShut {NoStop}%
\bibitem [{\citenamefont {Alicea}\ and\ \citenamefont
  {Fendley}(2016)}]{Aliceareview16}%
  \BibitemOpen
  \bibfield  {author} {\bibinfo {author} {\bibfnamefont {J.}~\bibnamefont
  {Alicea}}\ and\ \bibinfo {author} {\bibfnamefont {P.}~\bibnamefont
  {Fendley}},\ }\href {\doibase 10.1146/annurev-conmatphys-031115-011336}
  {\bibfield  {journal} {\bibinfo  {journal} {Annual Review of Condensed Matter
  Physics}\ }\textbf {\bibinfo {volume} {7}},\ \bibinfo {pages} {119} (\bibinfo
  {year} {2016})},\ \Eprint
  {http://arxiv.org/abs/https://doi.org/10.1146/annurev-conmatphys-031115-011336}
  {https://doi.org/10.1146/annurev-conmatphys-031115-011336} \BibitemShut
  {NoStop}%
\bibitem [{\citenamefont {Fendley}(2016)}]{Fendley2016}%
  \BibitemOpen
  \bibfield  {author} {\bibinfo {author} {\bibfnamefont {P.}~\bibnamefont
  {Fendley}},\ }\href {\doibase 10.1088/1751-8113/49/30/30lt01} {\bibfield
  {journal} {\bibinfo  {journal} {Journal of Physics A: Mathematical and
  Theoretical}\ }\textbf {\bibinfo {volume} {49}},\ \bibinfo {pages} {30LT01}
  (\bibinfo {year} {2016})}\BibitemShut {NoStop}%
\bibitem [{\citenamefont {Fendley}(2012)}]{Fendley2012}%
  \BibitemOpen
  \bibfield  {author} {\bibinfo {author} {\bibfnamefont {P.}~\bibnamefont
  {Fendley}},\ }\href {\doibase 10.1088/1742-5468/2012/11/p11020} {\bibfield
  {journal} {\bibinfo  {journal} {Journal of Statistical Mechanics: Theory and
  Experiment}\ }\textbf {\bibinfo {volume} {2012}},\ \bibinfo {pages} {P11020}
  (\bibinfo {year} {2012})}\BibitemShut {NoStop}%
\bibitem [{\citenamefont {Kitaev}(2006)}]{Kitaev06}%
  \BibitemOpen
  \bibfield  {author} {\bibinfo {author} {\bibfnamefont {A.}~\bibnamefont
  {Kitaev}},\ }\href {\doibase http://dx.doi.org/10.1016/j.aop.2005.10.005}
  {\bibfield  {journal} {\bibinfo  {journal} {Annals of Physics}\ }\textbf
  {\bibinfo {volume} {321}},\ \bibinfo {pages} {2 } (\bibinfo {year}
  {2006})}\BibitemShut {NoStop}%
\bibitem [{\citenamefont {Yao}\ and\ \citenamefont {Kivelson}(2007)}]{yao2007}%
  \BibitemOpen
  \bibfield  {author} {\bibinfo {author} {\bibfnamefont {H.}~\bibnamefont
  {Yao}}\ and\ \bibinfo {author} {\bibfnamefont {S.~A.}\ \bibnamefont
  {Kivelson}},\ }\href {\doibase 10.1103/PhysRevLett.99.247203} {\bibfield
  {journal} {\bibinfo  {journal} {Phys. Rev. Lett.}\ }\textbf {\bibinfo
  {volume} {99}},\ \bibinfo {pages} {247203} (\bibinfo {year}
  {2007})}\BibitemShut {NoStop}%
\bibitem [{\citenamefont {Fu}(2019)}]{Fu20191}%
  \BibitemOpen
  \bibfield  {author} {\bibinfo {author} {\bibfnamefont {J.}~\bibnamefont
  {Fu}},\ }\href {\doibase 10.1103/PhysRevB.100.195131} {\bibfield  {journal}
  {\bibinfo  {journal} {Phys. Rev. B}\ }\textbf {\bibinfo {volume} {100}},\
  \bibinfo {pages} {195131} (\bibinfo {year} {2019})}\BibitemShut {NoStop}%
\bibitem [{\citenamefont {Peri}\ \emph {et~al.}(2020)\citenamefont {Peri},
  \citenamefont {Ok}, \citenamefont {Tsirkin}, \citenamefont {Neupert},
  \citenamefont {Baskaran}, \citenamefont {Greiter}, \citenamefont {Moessner},\
  and\ \citenamefont {Thomale}}]{peri20}%
  \BibitemOpen
  \bibfield  {author} {\bibinfo {author} {\bibfnamefont {V.}~\bibnamefont
  {Peri}}, \bibinfo {author} {\bibfnamefont {S.}~\bibnamefont {Ok}}, \bibinfo
  {author} {\bibfnamefont {S.~S.}\ \bibnamefont {Tsirkin}}, \bibinfo {author}
  {\bibfnamefont {T.}~\bibnamefont {Neupert}}, \bibinfo {author} {\bibfnamefont
  {G.}~\bibnamefont {Baskaran}}, \bibinfo {author} {\bibfnamefont
  {M.}~\bibnamefont {Greiter}}, \bibinfo {author} {\bibfnamefont
  {R.}~\bibnamefont {Moessner}}, \ and\ \bibinfo {author} {\bibfnamefont
  {R.}~\bibnamefont {Thomale}},\ }\href {\doibase 10.1103/PhysRevB.101.041114}
  {\bibfield  {journal} {\bibinfo  {journal} {Phys. Rev. B}\ }\textbf {\bibinfo
  {volume} {101}},\ \bibinfo {pages} {041114} (\bibinfo {year}
  {2020})}\BibitemShut {NoStop}%
\bibitem [{\citenamefont {Wouters}\ \emph {et~al.}(2018)\citenamefont
  {Wouters}, \citenamefont {Katsura},\ and\ \citenamefont
  {Schuricht}}]{wouters18}%
  \BibitemOpen
  \bibfield  {author} {\bibinfo {author} {\bibfnamefont {J.}~\bibnamefont
  {Wouters}}, \bibinfo {author} {\bibfnamefont {H.}~\bibnamefont {Katsura}}, \
  and\ \bibinfo {author} {\bibfnamefont {D.}~\bibnamefont {Schuricht}},\ }\href
  {\doibase 10.1103/PhysRevB.98.155119} {\bibfield  {journal} {\bibinfo
  {journal} {Phys. Rev. B}\ }\textbf {\bibinfo {volume} {98}},\ \bibinfo
  {pages} {155119} (\bibinfo {year} {2018})}\BibitemShut {NoStop}%
\bibitem [{\citenamefont {Wada}\ \emph {et~al.}(2021)\citenamefont {Wada},
  \citenamefont {Sugimoto},\ and\ \citenamefont {Tohyama}}]{wada21}%
  \BibitemOpen
  \bibfield  {author} {\bibinfo {author} {\bibfnamefont {K.}~\bibnamefont
  {Wada}}, \bibinfo {author} {\bibfnamefont {T.}~\bibnamefont {Sugimoto}}, \
  and\ \bibinfo {author} {\bibfnamefont {T.}~\bibnamefont {Tohyama}},\ }\href
  {\doibase 10.1103/PhysRevB.104.075119} {\bibfield  {journal} {\bibinfo
  {journal} {Phys. Rev. B}\ }\textbf {\bibinfo {volume} {104}},\ \bibinfo
  {pages} {075119} (\bibinfo {year} {2021})}\BibitemShut {NoStop}%
\bibitem [{\citenamefont {Jordan}\ and\ \citenamefont
  {Wigner}(1928)}]{Jordan1928}%
  \BibitemOpen
  \bibfield  {author} {\bibinfo {author} {\bibfnamefont {P.}~\bibnamefont
  {Jordan}}\ and\ \bibinfo {author} {\bibfnamefont {E.}~\bibnamefont
  {Wigner}},\ }\href {\doibase 10.1007/BF01331938} {\bibfield  {journal}
  {\bibinfo  {journal} {Zeitschrift f{\"u}r Physik}\ }\textbf {\bibinfo
  {volume} {47}},\ \bibinfo {pages} {631} (\bibinfo {year} {1928})}\BibitemShut
  {NoStop}%
\bibitem [{\citenamefont {Lieb}\ \emph {et~al.}(1961)\citenamefont {Lieb},
  \citenamefont {Schultz},\ and\ \citenamefont {Mattis}}]{Lieb61}%
  \BibitemOpen
  \bibfield  {author} {\bibinfo {author} {\bibfnamefont {E.}~\bibnamefont
  {Lieb}}, \bibinfo {author} {\bibfnamefont {T.}~\bibnamefont {Schultz}}, \
  and\ \bibinfo {author} {\bibfnamefont {D.}~\bibnamefont {Mattis}},\ }\href
  {\doibase https://doi.org/10.1016/0003-4916(61)90115-4} {\bibfield  {journal}
  {\bibinfo  {journal} {Annals of Physics}\ }\textbf {\bibinfo {volume} {16}},\
  \bibinfo {pages} {407 } (\bibinfo {year} {1961})}\BibitemShut {NoStop}%
\bibitem [{\citenamefont {Sachdev}(2011)}]{sachdevbook}%
  \BibitemOpen
  \bibfield  {author} {\bibinfo {author} {\bibfnamefont {S.}~\bibnamefont
  {Sachdev}},\ }\href {\doibase 10.1017/CBO9780511973765} {\emph {\bibinfo
  {title} {Quantum Phase Transitions}}},\ \bibinfo {edition} {2nd}\ ed.\
  (\bibinfo  {publisher} {Cambridge University Press},\ \bibinfo {year}
  {2011})\BibitemShut {NoStop}%
\bibitem [{\citenamefont {McGinley}\ \emph {et~al.}(2017)\citenamefont
  {McGinley}, \citenamefont {Knolle},\ and\ \citenamefont
  {Nunnenkamp}}]{Mcginley2017}%
  \BibitemOpen
  \bibfield  {author} {\bibinfo {author} {\bibfnamefont {M.}~\bibnamefont
  {McGinley}}, \bibinfo {author} {\bibfnamefont {J.}~\bibnamefont {Knolle}}, \
  and\ \bibinfo {author} {\bibfnamefont {A.}~\bibnamefont {Nunnenkamp}},\
  }\href {\doibase 10.1103/PhysRevB.96.241113} {\bibfield  {journal} {\bibinfo
  {journal} {Phys. Rev. B}\ }\textbf {\bibinfo {volume} {96}},\ \bibinfo
  {pages} {241113} (\bibinfo {year} {2017})}\BibitemShut {NoStop}%
\bibitem [{\citenamefont {Ortiz}\ \emph {et~al.}(2014)\citenamefont {Ortiz},
  \citenamefont {Dukelsky}, \citenamefont {Cobanera}, \citenamefont {Esebbag},\
  and\ \citenamefont {Beenakker}}]{ort14}%
  \BibitemOpen
  \bibfield  {author} {\bibinfo {author} {\bibfnamefont {G.}~\bibnamefont
  {Ortiz}}, \bibinfo {author} {\bibfnamefont {J.}~\bibnamefont {Dukelsky}},
  \bibinfo {author} {\bibfnamefont {E.}~\bibnamefont {Cobanera}}, \bibinfo
  {author} {\bibfnamefont {C.}~\bibnamefont {Esebbag}}, \ and\ \bibinfo
  {author} {\bibfnamefont {C.}~\bibnamefont {Beenakker}},\ }\href {\doibase
  10.1103/PhysRevLett.113.267002} {\bibfield  {journal} {\bibinfo  {journal}
  {Phys. Rev. Lett.}\ }\textbf {\bibinfo {volume} {113}},\ \bibinfo {pages}
  {267002} (\bibinfo {year} {2014})}\BibitemShut {NoStop}%
\bibitem [{\citenamefont {Sau}\ \emph {et~al.}(2011)\citenamefont {Sau},
  \citenamefont {Halperin}, \citenamefont {Flensberg},\ and\ \citenamefont
  {Das~Sarma}}]{sau11}%
  \BibitemOpen
  \bibfield  {author} {\bibinfo {author} {\bibfnamefont {J.~D.}\ \bibnamefont
  {Sau}}, \bibinfo {author} {\bibfnamefont {B.~I.}\ \bibnamefont {Halperin}},
  \bibinfo {author} {\bibfnamefont {K.}~\bibnamefont {Flensberg}}, \ and\
  \bibinfo {author} {\bibfnamefont {S.}~\bibnamefont {Das~Sarma}},\ }\href
  {\doibase 10.1103/PhysRevB.84.144509} {\bibfield  {journal} {\bibinfo
  {journal} {Phys. Rev. B}\ }\textbf {\bibinfo {volume} {84}},\ \bibinfo
  {pages} {144509} (\bibinfo {year} {2011})}\BibitemShut {NoStop}%
\bibitem [{\citenamefont {Fidkowski}\ \emph {et~al.}(2012)\citenamefont
  {Fidkowski}, \citenamefont {Alicea}, \citenamefont {Lindner}, \citenamefont
  {Lutchyn},\ and\ \citenamefont {Fisher}}]{Fid12}%
  \BibitemOpen
  \bibfield  {author} {\bibinfo {author} {\bibfnamefont {L.}~\bibnamefont
  {Fidkowski}}, \bibinfo {author} {\bibfnamefont {J.}~\bibnamefont {Alicea}},
  \bibinfo {author} {\bibfnamefont {N.~H.}\ \bibnamefont {Lindner}}, \bibinfo
  {author} {\bibfnamefont {R.~M.}\ \bibnamefont {Lutchyn}}, \ and\ \bibinfo
  {author} {\bibfnamefont {M.~P.~A.}\ \bibnamefont {Fisher}},\ }\href {\doibase
  10.1103/PhysRevB.85.245121} {\bibfield  {journal} {\bibinfo  {journal} {Phys.
  Rev. B}\ }\textbf {\bibinfo {volume} {85}},\ \bibinfo {pages} {245121}
  (\bibinfo {year} {2012})}\BibitemShut {NoStop}%
\bibitem [{\citenamefont {Moudgalya}\ and\ \citenamefont
  {Motrunich}(2022)}]{moudgalya22}%
  \BibitemOpen
  \bibfield  {author} {\bibinfo {author} {\bibfnamefont {S.}~\bibnamefont
  {Moudgalya}}\ and\ \bibinfo {author} {\bibfnamefont {O.~I.}\ \bibnamefont
  {Motrunich}},\ }\href {\doibase 10.1103/PhysRevX.12.011050} {\bibfield
  {journal} {\bibinfo  {journal} {Phys. Rev. X}\ }\textbf {\bibinfo {volume}
  {12}},\ \bibinfo {pages} {011050} (\bibinfo {year} {2022})}\BibitemShut
  {NoStop}%
\bibitem [{\citenamefont {Su}\ \emph {et~al.}(1979)\citenamefont {Su},
  \citenamefont {Schrieffer},\ and\ \citenamefont {Heeger}}]{Su791}%
  \BibitemOpen
  \bibfield  {author} {\bibinfo {author} {\bibfnamefont {W.~P.}\ \bibnamefont
  {Su}}, \bibinfo {author} {\bibfnamefont {J.~R.}\ \bibnamefont {Schrieffer}},
  \ and\ \bibinfo {author} {\bibfnamefont {A.~J.}\ \bibnamefont {Heeger}},\
  }\href {\doibase 10.1103/PhysRevLett.42.1698} {\bibfield  {journal} {\bibinfo
   {journal} {Phys. Rev. Lett.}\ }\textbf {\bibinfo {volume} {42}},\ \bibinfo
  {pages} {1698} (\bibinfo {year} {1979})}\BibitemShut {NoStop}%
\bibitem [{\citenamefont {Berezin}\ and\ \citenamefont
  {Marinov}(1975)}]{berezin1975}%
  \BibitemOpen
  \bibfield  {author} {\bibinfo {author} {\bibfnamefont {F.}~\bibnamefont
  {Berezin}}\ and\ \bibinfo {author} {\bibfnamefont {M.}~\bibnamefont
  {Marinov}},\ }\href
  {http://www.jetpletters.ac.ru/ps/1476/article_22521.shtml} {\bibfield
  {journal} {\bibinfo  {journal} {Sov. Phys. JETP Lett.}\ }\textbf {\bibinfo
  {volume} {21}},\ \bibinfo {pages} {320} (\bibinfo {year} {1975})}\BibitemShut
  {NoStop}%
\bibitem [{\citenamefont {Fu}\ \emph {et~al.}(2018)\citenamefont {Fu},
  \citenamefont {Knolle},\ and\ \citenamefont {Perkins}}]{Fu20181}%
  \BibitemOpen
  \bibfield  {author} {\bibinfo {author} {\bibfnamefont {J.}~\bibnamefont
  {Fu}}, \bibinfo {author} {\bibfnamefont {J.}~\bibnamefont {Knolle}}, \ and\
  \bibinfo {author} {\bibfnamefont {N.~B.}\ \bibnamefont {Perkins}},\ }\href
  {\doibase 10.1103/PhysRevB.97.115142} {\bibfield  {journal} {\bibinfo
  {journal} {Phys. Rev. B}\ }\textbf {\bibinfo {volume} {97}},\ \bibinfo
  {pages} {115142} (\bibinfo {year} {2018})}\BibitemShut {NoStop}%
\bibitem [{\citenamefont {Fu}(2018)}]{Fu20182}%
  \BibitemOpen
  \bibfield  {author} {\bibinfo {author} {\bibfnamefont {J.}~\bibnamefont
  {Fu}},\ }\href {\doibase 10.1103/PhysRevB.98.214432} {\bibfield  {journal}
  {\bibinfo  {journal} {Phys. Rev. B}\ }\textbf {\bibinfo {volume} {98}},\
  \bibinfo {pages} {214432} (\bibinfo {year} {2018})}\BibitemShut {NoStop}%
\bibitem [{\citenamefont {Feng}\ \emph {et~al.}(2007)\citenamefont {Feng},
  \citenamefont {Zhang},\ and\ \citenamefont {Xiang}}]{Feng07}%
  \BibitemOpen
  \bibfield  {author} {\bibinfo {author} {\bibfnamefont {X.-Y.}\ \bibnamefont
  {Feng}}, \bibinfo {author} {\bibfnamefont {G.-M.}\ \bibnamefont {Zhang}}, \
  and\ \bibinfo {author} {\bibfnamefont {T.}~\bibnamefont {Xiang}},\ }\href
  {\doibase 10.1103/PhysRevLett.98.087204} {\bibfield  {journal} {\bibinfo
  {journal} {Phys. Rev. Lett.}\ }\textbf {\bibinfo {volume} {98}},\ \bibinfo
  {pages} {087204} (\bibinfo {year} {2007})}\BibitemShut {NoStop}%
\bibitem [{\citenamefont {Schmidt}\ \emph {et~al.}(2008)\citenamefont
  {Schmidt}, \citenamefont {Dusuel},\ and\ \citenamefont {Vidal}}]{schmidt08}%
  \BibitemOpen
  \bibfield  {author} {\bibinfo {author} {\bibfnamefont {K.~P.}\ \bibnamefont
  {Schmidt}}, \bibinfo {author} {\bibfnamefont {S.}~\bibnamefont {Dusuel}}, \
  and\ \bibinfo {author} {\bibfnamefont {J.}~\bibnamefont {Vidal}},\ }\href
  {\doibase 10.1103/PhysRevLett.100.057208} {\bibfield  {journal} {\bibinfo
  {journal} {Phys. Rev. Lett.}\ }\textbf {\bibinfo {volume} {100}},\ \bibinfo
  {pages} {057208} (\bibinfo {year} {2008})}\BibitemShut {NoStop}%
\bibitem [{\citenamefont {Lieb}(1994)}]{lieb94}%
  \BibitemOpen
  \bibfield  {author} {\bibinfo {author} {\bibfnamefont {E.~H.}\ \bibnamefont
  {Lieb}},\ }\href {\doibase 10.1103/PhysRevLett.73.2158} {\bibfield  {journal}
  {\bibinfo  {journal} {Phys. Rev. Lett.}\ }\textbf {\bibinfo {volume} {73}},\
  \bibinfo {pages} {2158} (\bibinfo {year} {1994})}\BibitemShut {NoStop}%
\bibitem [{\citenamefont {Niu}\ \emph {et~al.}(2012)\citenamefont {Niu},
  \citenamefont {Chung}, \citenamefont {Hsu}, \citenamefont {Mandal},
  \citenamefont {Raghu},\ and\ \citenamefont {Chakravarty}}]{Niu12}%
  \BibitemOpen
  \bibfield  {author} {\bibinfo {author} {\bibfnamefont {Y.}~\bibnamefont
  {Niu}}, \bibinfo {author} {\bibfnamefont {S.~B.}\ \bibnamefont {Chung}},
  \bibinfo {author} {\bibfnamefont {C.-H.}\ \bibnamefont {Hsu}}, \bibinfo
  {author} {\bibfnamefont {I.}~\bibnamefont {Mandal}}, \bibinfo {author}
  {\bibfnamefont {S.}~\bibnamefont {Raghu}}, \ and\ \bibinfo {author}
  {\bibfnamefont {S.}~\bibnamefont {Chakravarty}},\ }\href {\doibase
  10.1103/PhysRevB.85.035110} {\bibfield  {journal} {\bibinfo  {journal} {Phys.
  Rev. B}\ }\textbf {\bibinfo {volume} {85}},\ \bibinfo {pages} {035110}
  (\bibinfo {year} {2012})}\BibitemShut {NoStop}%
\bibitem [{\citenamefont {Greiter}\ \emph {et~al.}(2014)\citenamefont
  {Greiter}, \citenamefont {Schnells},\ and\ \citenamefont
  {Thomale}}]{Greiter14}%
  \BibitemOpen
  \bibfield  {author} {\bibinfo {author} {\bibfnamefont {M.}~\bibnamefont
  {Greiter}}, \bibinfo {author} {\bibfnamefont {V.}~\bibnamefont {Schnells}}, \
  and\ \bibinfo {author} {\bibfnamefont {R.}~\bibnamefont {Thomale}},\ }\href
  {\doibase https://doi.org/10.1016/j.aop.2014.08.013} {\bibfield  {journal}
  {\bibinfo  {journal} {Annals of Physics}\ }\textbf {\bibinfo {volume}
  {351}},\ \bibinfo {pages} {1026} (\bibinfo {year} {2014})}\BibitemShut
  {NoStop}%
\end{thebibliography}%

\end{document}